\documentclass[
    ,final            
  ]
  {aipproc}

\layoutstyle{6x9}

\begin{document}

\title{News From The $\gamma$~Cephei Planetary System}


\author{Michael Endl}{
  address={McDonald Observatory, University of Texas at Austin, TX, 78712, USA}
}

\author{William D. Cochran}{
  address={McDonald Observatory, University of Texas at Austin, TX, 78712, USA}
}

\author{Artie P. Hatzes}{
  address={Th\"uringer Landessternwarte Tautenburg, Germany}
}

\author{Robert A. Wittenmyer}{
  address={Department of Astrophysics, University of New South Wales, Sydney, Australia}
}

\begin{abstract}
The $\gamma$~Cephei planetary system is one of the most interesting systems due to several reasons:
1.) it is the first planet candidate detected by precise radial velocity (RV) measurements that was discussed
in the literature (Campbell et al.\,1988); 2.) it is a tight binary system with $a \approx 20$~AU; and 3.) the
planet host star is an evolved K-type star. In Hatzes et al.~(2003) we confirmed the presence of the planetary
companion with a minimum mass of 1.7~M$_{\rm Jup}$ at 2~AU. In this paper we present additional eight years of precise RV data from
the Harlan J. Smith 2.7~m Telescope and its Tull C\'oude spectrograph at McDonald Observatory. The 900~d signal, that
is interpreted as the presence of the giant planetary companion, is strongly confirmed by adding the new data. 
We present an updated orbital solution for the planet, which shows that the planet is slightly more massive and
the orbit more circular than previous results have suggested.
An intensive high-cadence week of RV observations in 2007 revealed that $\gamma$~Cep~A is a multi-periodic pulsator. 
We discuss this issue within the context of searching for additional planets in this system.  
\end{abstract}

\maketitle


\section{Introduction}

$\gamma$~Cep is a nearby ($d=13.8$~pc), bright ($V=3.2$) K1 IV-III star, that has the distinction to be the very first
candidate for an extrasolar planetary companion detected by precise RV measurements that was discussed in the literature. The pioneering
RV survey at the CFHT carried out by Bruce Campbell and Gordon Walker discovered that $\gamma$~Cep is not only a single-lined
binary, but also that it shows periodic low-amplitude RV variations that might indicate the presence of a giant planetary companion with
a period around 900~days (Campbell et al.~1988). However, the same team later cast doubt on the existence of this planet 
(Walker et al.~1992). They argued that this signal could be caused by intrinsic variability of the star itself. More 
recently, Hatzes et al.~(2003) used over
14 years of precise RV data obtained by the McDonald Observatory planet search to demonstrate that the 900~d signal stayed coherent in phase
and amplitude of the entire time span. We also showed that there is no significant variability in the chromospheric emission level and
line bisectors based on the McDonald spectra of $\gamma$~Cep. We concluded that the 900~d signal is indeed due to a planetary companion with a
minimum mass of 1.7~M$_{\rm Jup}$ orbiting at $a=2.1$~AU. 

The detection of this planet in a relatively tight binary system with $a \approx 20$~AU lead to a host of follow-up investigations ranging from 
dynamical studies (e.g. Dvorak et al.~2003, Haghighipour~2006), to planet formation modeling (e.g. Kley \& Nelson~2008, Jang-Condell et al.~2008), to 
additional work combining 
existing RV data with astrometric measurements (Torres 2007), and to direct imaging campaigns (Neuh\"auser et al.~2007). In particular the 
combination of the last two papers significantly improved the parameters of the $\gamma$~Cep stellar binary system. The orbital period of the binary is
$67\pm1.4$ years with an eccentricity of $0.41$ and a semi-major axis of $20.2\pm0.7$~AU. The primary has a mass of M$_{\rm A} = 1.4\pm0.12$~M$_{\odot}$ and
the secondary is an M dwarf with M$_{\rm B} = 0.41 \pm 0.02$~M$_{\odot}$.

In this paper we present our results based on additional eight years of precise RV measurements from the planet search program at the Harlan J. Smith 2.7\,m telescope 
(HJST) and its Tull C\'oude spectrograph at McDonald Observatory.    

\section{Updated Orbital and Planetary Parameters}

\begin{figure}[t]
  \includegraphics[height=.83\textwidth, angle=270]{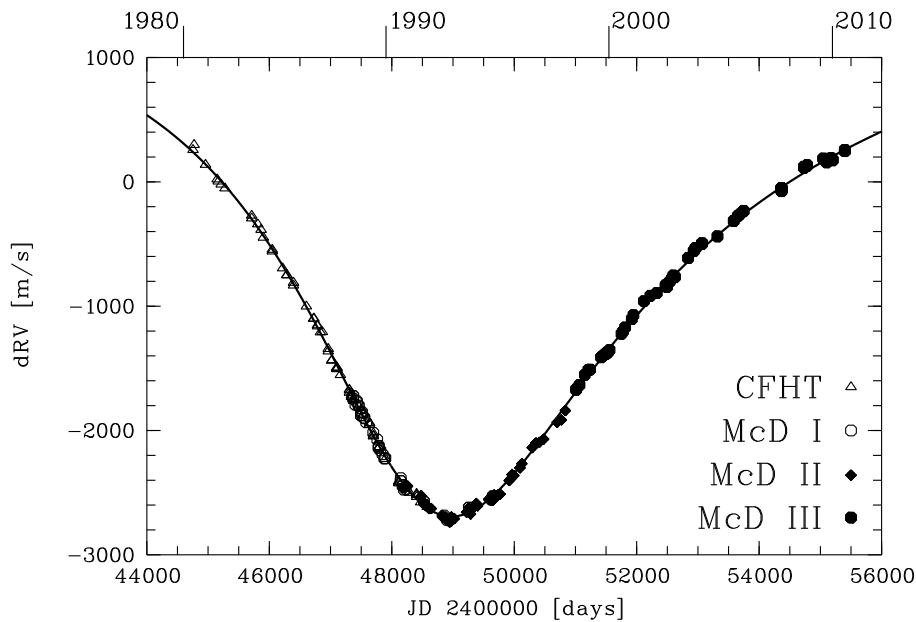}
  \caption{Almost 30 years of precise RV data for $\gamma$~Cep~A. The four different
data sets are from the CFHT (triangles) and from the 3 different phases of the McDonald
Observatory planet search (circles, diamonds, filled circles). The solid line displays
the RV orbit due to the stellar secondary.}
\end{figure}

After the announcement of the planet in 2003, we have continued to observe $\gamma$~Cep as part of the long-term precise Doppler survey
at the HJST. We are currently in phase III of the program that utilizes the entire available spectral bandwidth of the iodine cell that
we use to measure precise differential RVs. Phase I and II used either different reference lines (e.g. telluric lines) or 
instrumental configurations. A detailed description of the different phases of our program can be found in Hatzes et al.~(2003).   
Figure 1 displays the entire RV data starting with the CFHT results and ending with the McD phase III. These data cover the periastron passage of the binary 
orbit (solid line). With these new data in hand, we can confirm or falsify the planetary hypothesis by testing if the residual signal stayed
constant in period, phase and amplitude. 

Fig.\,2 shows the Lomb-Scargle periodogram of the RV residuals after subtracting the large binary motion. The 900~d signal is clearly present in our data and
its statistical significance is higher than in 2003. The false-alarm-probability of the peak is $\approx 10^{-30}$. Our phase III spectra contain the
Ca II H\&K lines that can be used as proxies for magnetic activity of the star. We determine a Ca line S-index that corresponds to the level of chromospheric emission 
in the Ca H \& K line cores. If any correlation between the S-indices and the RV data is found it would cast serious doubt on the existence of a companion. 
Fig.\,3 contains the periodogram of our S-index measurements. The vertical dashed lines indicates the 900~d period for the planet. It shows that there is no
isolated peak at the orbital period, further strengthening the case for the planet. However, there is a peak of moderate power at a significantly shorter period of
$\approx 800$~d. Although not highly significant, this peak could indicate the rotational period of the star. Our re-analysis of the CFHT FWHM measurements in
Hatzes et al. also revealed a shorter period that can be the stellar rotational period. 

We conclude that the addition of the new data strongly confirms the presence of the 900~d signal residual signal and 
that there appears to be no connection between this signal and intrinsic variability of the host star.      
 
\begin{figure}
  \includegraphics[height=.83\textwidth, angle=270]{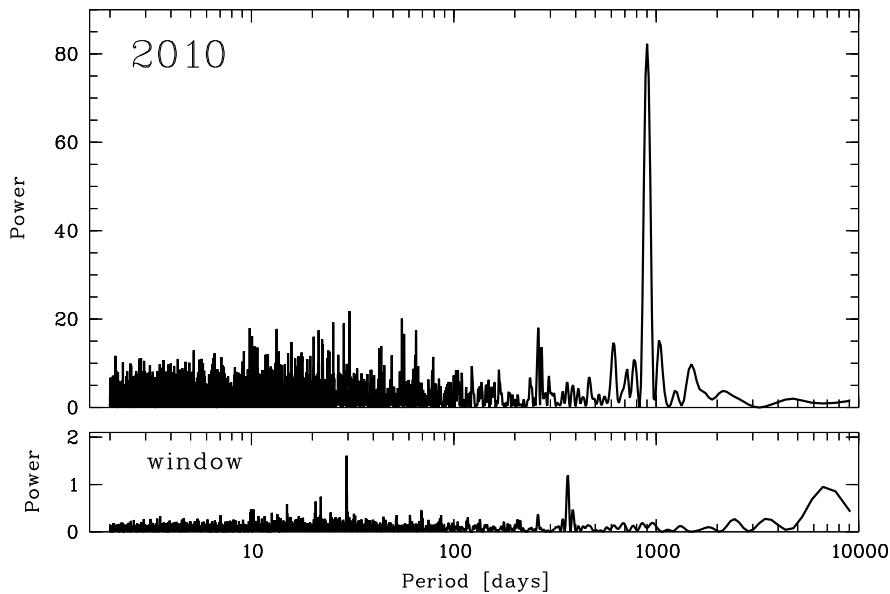}
  \caption{Lomb-Scargle periodogram of the RV residuals after subtracting the binary
orbit (upper panel) and the window function of our observations (lower panel). 
The peak at 900~days has an estimated false-alarm-probability of 10$^{-30}$.}
\end{figure}

We derived a simultaneous Keplerian orbital solution for both the binary and the planetary orbit using {\it Gaussfit} (Jefferys et al. 1988). 
The RV offsets between the four individual data sets were included as fit parameters.
The only parameter that we held fixed is the period of the binary. We used 24650 days, the current best value following Torres (2007) and Neuh\"auser et al.~(2007). 
Letting this parameter vary freely lead to shorter binary periods, which are inconsistent with the global solution based on historic low precision RVs and imaging
results. Despite that the 30~years of precise RVs cover the periastron passage and almost half of the entire orbit, it is not sufficient to 
derive the correct binary period using these data alone. The remaining orbital parameters for the binary are in good agreement with the published values: 
K$=1932\pm3$\,m\,s$^{-1}$, $e=0.41\pm0.001$, $\omega=162\pm0.3$ and T$_{\rm Per}=2448502\pm8.5$~days. The updated parameters for the 
planet are given in Table~1. The biggest difference to the previously published values are a higher minimum mass of $1.85\pm0.16$\,M$_{\rm Jup}$ (previously
$1.7\pm0.4$\,M$_{\rm Jup}$) and a lower eccentricity of $0.049\pm0.034$ (previously $0.12\pm0.05$). Fig.\,4 shows the 4 different 
RV data sets (after subtracting the binary motion) along with the planetary orbit phased to the 900~d period.        

\begin{table}[b]
\begin{tabular}{lrrr}
\hline
\tablehead{1}{r}{b}{parameter} & 
\tablehead{1}{r}{b}{value} & 
\tablehead{1}{r}{b}{error} & 
\tablehead{1}{r}{b}{units} \\
\hline
Period & 903.3 & 1.5 & [days]   \\
K & 31.1 & 0.97 & [m\,s$^{-1}$] \\
e & 0.049 & 0.034 & \\
T$_{\rm per}$ & 2453227 & 87 & [days] \\
$\omega$ & 94.6 & 34.6 & [deg] \\
m $\sin i$ & 1.85 & 0.16 & [M$_{\rm Jup}$]\\
a & 2.05 & 0.06 & [AU]\\ 
\hline
\end{tabular}
\caption{New orbital and planetary parameters for $\gamma$~Cep~Ab}
\label{tab:a}
\end{table}

\begin{figure}
  \includegraphics[height=.83\textwidth, angle=270]{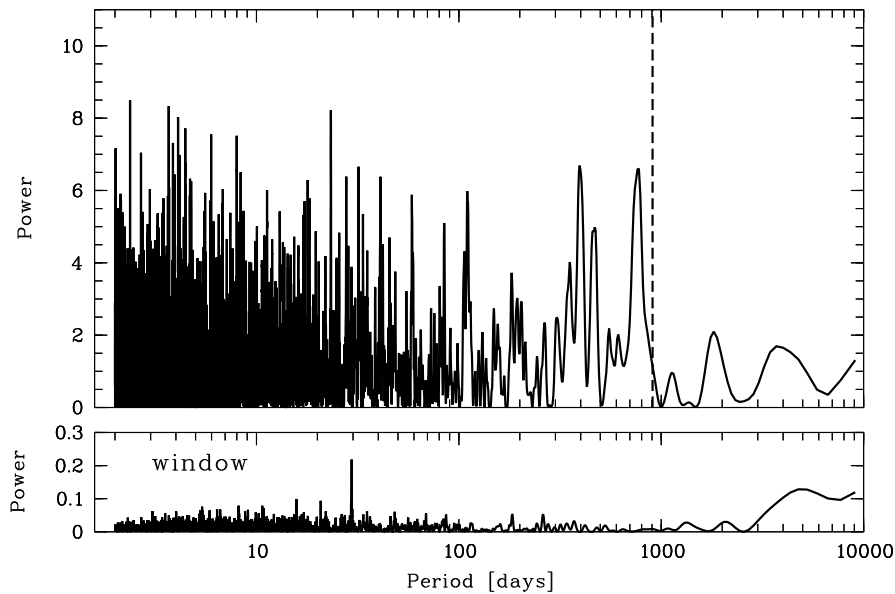}
  \caption{Lomb-Scargle periodogram of the Ca II H\& K line indices from our 
McDonald spectra (upper panel) and their window function (lower panel). The vertical
dashed line displays the period of the planetary companion. There is some power in a
neighboring peak at $\approx 800$~days that might indicate the rotational period of 
$\gamma$~Cep~A.}
\end{figure}

\clearpage

\begin{figure}
  \includegraphics[height=.85\textwidth, angle=270]{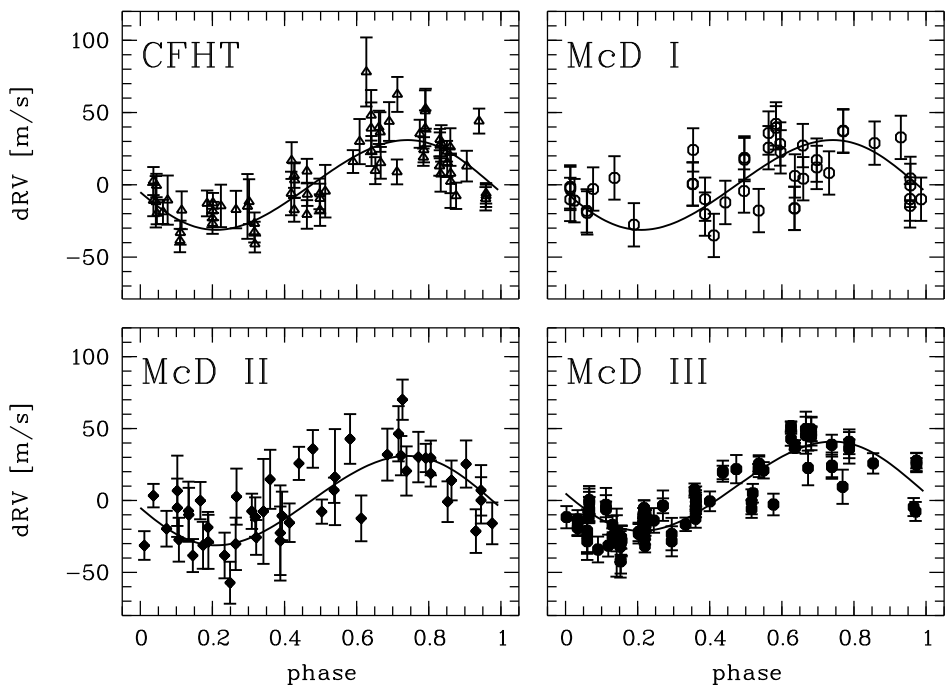}
  \caption{The four RV data sets phased to the orbital period (903~d) of the planetary companion.
The residuals scatter around this new orbital solution are: CFHT: 15.4\,m\,s$^{-1}$, McD I: 19.5\,m\,s$^{-1}$,
McD II: 17.7\,m\,s$^{-1}$, and McD III: 7.7\,m\,s$^{-1}$}.
\end{figure}

\subsection{Searching For Additional Planets}

The residual RV scatter of $\approx 8$\,m\,s$^{-1}$ is more than twice as large as the internal measurement
uncertainties of 3\,m\,s$^{-1}$ for the McDonald phase III data. This excess scatter could be a sign of
additional, previously undetected planets in the system. Based on dynamical simulations, Haghighipour~(2006) showed that
low-mass planets with $a\le0.8$~AU should remain stable in this system.  
However, a period search in the residuals to our binary plus single planet fit did not (yet) reveal any significant signals (see
Fig.\,5). 

\begin{figure}
  \includegraphics[height=.85\textwidth, angle=270]{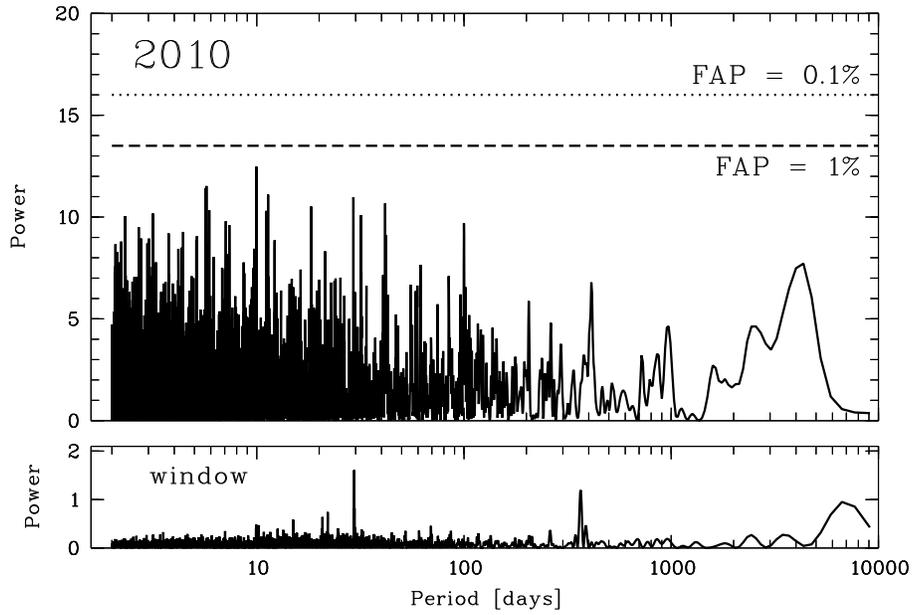}
  \caption{Lomb-Scargle periodogram of the RV residuals after subtracting the binary and the single
planet model. All peaks have a false-alarm-probability (FAP) of $>1\%$. Therefore, we have no evidence 
for additional RV signals in our data.}
\end{figure}

In September 2007 we used seven consecutive nights at the HJST to perform an intensive asteroseismological study of
$\gamma$~Cep~A. We collected over 1200 individual RV measurements and discovered that the star is a low-amplitude
multi-periodic pulsator. The results of this observing run are described in more detail in Endl et al.~(2009). We identified
12 different frequencies with amplitudes ranging from 1.5 to 4.1\,m\,s$^{-1}$. A section of one night from this campaign is
displayed in Fig.\,6. The oscillation spectrum of $\gamma$~Cep~A has a
large spacing of 15\,$\mu$Hz, which is proportional to the mass and radius of the star. Using a mass of 1.4\,M$_{\odot}$ we 
derive a radius of 4.9~R$_{\odot}$. This value is right in between the two interferometrically measured radius values of 
$4.79 \pm 0.06$~R$_{\odot}$ (Nordgren et al.~1999) and $5.01 \pm 0.05$~R$_{\odot}$ (Baines et al.~2009).  

\begin{figure}
  \includegraphics[height=.85\textwidth, angle=270]{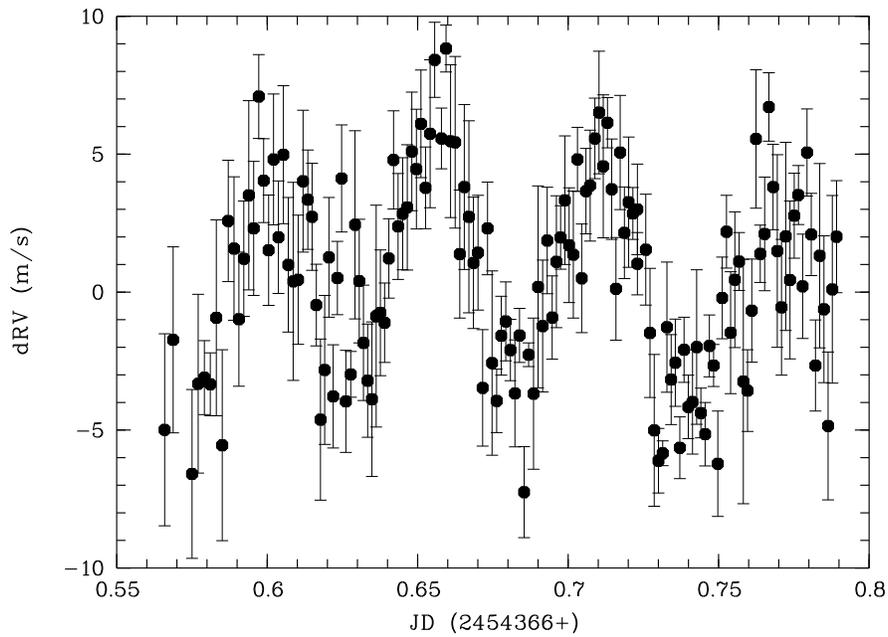}
  \caption{A small part of the high-cadence asteroseismology RV data we have obtained in 2007. Stellar 
oscillations with amplitudes of a few m\,s$^{-1}$ are clearly visible in these data.}
\end{figure}

The oscillations of $\gamma$~Cep~A are the cause of the observed excess scatter. Although we have currently no evidence for additional
planets in the system, such a search is limited by the short-term variability due to the pulsations. To account for this additional
RV signal in our data, we have increased the error bars on our phase III data artificially by a factor of two before performing the orbital fits 
presented in this paper. Otherwise, the small formal errors of the phase III data, which do not take the oscillations into account, 
would have severely underestimated the true uncertainties of these data.    

The pulsation signal of a few m\,s$^{-1}$ is present in the entire RV data set and it limits our ability to search for lower-mass planets
that would produce RV signals with similar amplitudes. The exposures times for a bright star like $\gamma$~Cep are short enough so that
we sampled these pulsations at random phases, rather than average them out. Therefore, as the next step, we plan to define and subtract the 
oscillation signal from the entire 30 years of RV data. This should allow us to reach the highest sensitivity for additional planets in the 
$\gamma$~Cep system.    

\begin{theacknowledgments}
We would like to thank the many observers of the McDonald Observatory planet search program at the HJST that have 
obtained spectra of $\gamma$~Cep over the years: Phillip J. MacQueen, Stuart I. Barnes, Diane Paulson, Paul Robsertson, Erik Brugamyer and Candace Gray.
Jacob Bean and B\'arbara Castanheira helped with the seismology campaign on this star. We are 
grateful to Bruce Campbell, Gordon A. H. Walker and Stephenson Yang for their CFHT data. M.E. and
W.D.C acknowledge support by the National Aeronautics and Space Administration under Grants NNX07AL70G and 
NNX09AB30G issued through the Origins of Solar System Program.   
\end{theacknowledgments}



\begin{thebibliography}{9}

\bibitem{Bain:2009}
E.~K.~Baines, H.~A.~McAlister, T.~A.~ten Brummelaar, J.~Sturmann, L.~Sturmann, N.~H.~Turner, S.~T,~Ridgway,
\emph{ApJ}, \textbf{701}, 154--162 (2009).

\bibitem{Camp:1988}
B.~Campbell, G.~A.~H.~Walker, and S.~Yang, \emph{ApJ}, \textbf{331}, 902--921 (1988).

\bibitem{Dvor:2003}
R.~Dvorak, E.~Pilat-Lohinger, B.~Funk, F.~Freistetter, \emph{A\&A}, \textbf{398}, 
L1--L4 (2003). 

\bibitem{Endl:2009}
M.~Endl, J.~L.~Bean, R.~A.~Wittenmyer, A.~P.~Hatzes, B.~G.~Castanheira, W.~D.~Cochran, 
AIP Conference Proceedings 1170, 543--544 (2009). 

\bibitem{Hagh:2006}
N.~Haghighipour,  \emph{ApJ}, \textbf{644}, 543--550 (2006).

\bibitem{Hatz:2003}
A.~P.~Hatzes, W.~D.~Cochran, M.~Endl, B.~McArthur, B.~Paulson, G.~A.~H.~Walker, B.~Campbell, and 
S.~Yang, \emph{ApJ}, \textbf{599}, 1383--1394 (2003).

\bibitem{Kley:2008}
W.~Kley, and R.~P.~Nelson, \emph{A\&A}, \textbf{486}, 617--628 (2008).

\bibitem{Jang:2008}
H.~Jang-Condell, M.~Mugrauer, and T.~Schmidt,\emph{ApJ}, \textbf{683}, L191--L194 (2008). 

\bibitem{Jeff:1988}
W.~H.~Jefferys, M.~J.~Fitzpatrick, B.~E.~McArthur, \emph{CeMec}, \textbf{41}, 39--49 (1988). 

\bibitem{Neuh:2007}
R.~Neuh\"auser, M.~Mugrauer, M.~Fukagawa, G.~Torres, T.~Schmidt, \emph{A\&A}, \textbf{462}, 777--780 (2007).

\bibitem{Nord:1999}
T.~E.~Nordgren, M.~E.~Germain, J.~A.~Benson, D.~Mozurkewich, J.~J.~Sudol, N.~M.~II~Elias, A.~R.~Hajian, N.~M.~White, D.~J.~Hutter, K.~J.~Johnston,
F.~S.~Gauss, J.~T.~Armstrong, T.~A.~Pauls, L.~J.~Rickard, \emph{AJ}, \textbf{118}, 3032--3038 (1999).

\bibitem{Torr:2007}
G.~Torres, \emph{ApJ}, \textbf{654}, 1095--1109 (2007).

\bibitem{Walk:1992}
G.~A.~H.~Walker, D.~A.~Bohlender, A.~R.~Walker, A.~W.~Irwin, S.~Yang and A.~Larson, \emph{ApJ},
\textbf{396}, L91--L94 (1992).

\end{thebibliography}
\end{document}